\documentclass[prl,aps,floatfix,nofootinbib,twocolumn,letterpaper,reprint]{revtex4}
\usepackage{graphicx}
\usepackage{amssymb} 
\usepackage{amsmath} 
\usepackage{amsfonts} 
\usepackage{amsbsy}
\usepackage{color}
\def\lb{\langle} 
\def\rb{\rangle} 
 
\def\be{\begin{equation}} 
\def\ee{\end{equation}}

\def\kr{$^{96}$Kr}
\def\sr{$^{96}$Sr}
\def\zr{$^{96}$Zr}
\def\ba{$^{140}$Ba}
\def\xe{$^{140}$Xe}
\def\te{$^{140}$Te}

\begin{document} 

\title{Angular Momentum of Fission Fragments
}
\author{G.F.~Bertsch$^{1}$, T.~Kawano$^{2}$ and L.M.~Robledo$^{3}$}
 
\affiliation{$^{1}$Department of Physics and Institute of Nuclear Theory, 
University of Washington, Seattle, Washington 98915, USA\\ 
$^{2}$Theoretical Division, Los Alamos National Laboratory, 
Los Alamos, New Mexico 87545, USA\\
Departamento de F\'{i}sica Te\'{o}rica, Universidad Aut\'{o}noma de Madrid,
28049-Madrid, Spain
}
 
\begin{abstract}
The strong deformation present immediately
after scission has consequences for the angular momentum population
of the fragments as well as the angular distribution of their
decay radiation.  We find that the usual spin-cutoff parameterization 
describes very well the 
angular momentum distribution 
associated with the deformation of the  fragments 
at the scission point.  Depending on the deformation, its contribution can
be comparable to the thermal contribution to the angular momentum of the
newly formed fragments.
The $M$-distribution of the angular
momentum is highly polarized and gives rise to large anisotropies
in the subsequent gamma cascade.  We treat
in detail a typical gamma  cascade in a daughter nucleus,
following usual model assumptions except for the anisotropy of
the initial state.
In principle, the observed anisotropy can 
provide information on the relative amounts of deformation 
and thermal energy present at the scission point.  
\end{abstract}
\maketitle 
 
{\it Introduction.}
Fission is a very complex nuclear reaction, both
before and after the scission point.  In principle the post-scission
theory should be simpler because one is treating ordinary decay
processes (gamma and neutron emission) in nearly isolated mid-mass
nuclei.  However, there are differences from the decays of the
excited nuclei produced in compound-nucleus reactions.  Most importantly,
the fission fragments may start in a state of high deformation aligned
along the fission axis.  Until recently \cite{bo07},
the angular momentum of the fragments as been treated
statistically, ignoring the specific consequences of the 
deformation.  Without inclusion of deformation effects,
statistical modeling fails to reproduce average angular momenta by
as much as a factor of two \cite{st14}.  

The goal of this article is 
to calculate the effects of the deformation on the angular momentum
distribution of the nascent fission fragments and their subsequent
decay.  The
theory of the scission process is now under active development and many
details are still obscure.  One promising approach to determine properties
of the fragments immediately following scission is time-dependent density 
functional theory \cite{bu15} (see also Ref. \cite{be18}).
The relation between deformation and angular
momentum content can be reasonably modeled in well-established mean-field
theories such as  Hartree-Fock (HF) or Hartree-Fock-Bogoliubov (HFB).
In the first section below, we estimate the average angular momentum
in deformed configurations using
one of the popular energy density functionals (Gogny D1S).  In the
section after that we determine the angular momentum 
probability distribution by projection.
We find that the shape of the $J$ distribution coming
from deformation is nearly identical to the shape assumed in the statistical
theory, differing only by the parameter controlling the average
width $\langle J^2\rangle$.

An interesting observable that hardly been used in the past is the
angular distribution of the decay gamma rays.  Due to the
alignment of the deformed fragment along the fission axis,
the $M$ distribution of its states favors $M=0$ along the axis.
Here we examine the angular distributions modeling 
the cascade from a fully aligned initial population.
We find that the effect on the gamma  distribution can approach
a factor of two in anisotropy.  This effect is certainly
measurable even in the presence of a large contribution from
the isotropic quasi-particle contribution. The sign of the anisotropy
is opposite for dipole and quadrupole photons, and in fact
the both kinds of anisotropy have been seen in the final 
decays to the ground states of daughter nuclei \cite{wi72}.

Of course the angular momentum distribution 
also has contributions from quasi-particle excitations.  Typically the pre-scission
state is already highly excited above the collective potential energy
surface and that excitation energy will be carried over to the
post-scission fragments.  The strong alignment of the deformed initial
configuration is degraded by the presence of quasiparticles and both
$\langle J^2 \rangle$ and the angular distributions will be affected.
However, in view of all of the uncertainties in the
present theory at the scission point we have not attempted to make
a quantitative estimate of the resulting cascade angular distributions.

{\it Angular momentum of aligned deformed nuclei.}
The
deformation and alignment of the fission fragments requires that the 
wave function be a coherent superposition of
angular momentum states.  To determine the angular momentum content,
we take the wave function from self-consistent mean-field theory.
The first question
we address is the relationship between deformation as characterized by
the Bohr parameter $\beta$ and the average squared angular momentum $\langle
J^2 \rangle$.   The second question is how the angular momentum is
distributed, ie. the probability distribution $P(J)$ given its average
for the configuration.  
These relationships were also studied in Ref. \cite{bo07} using
different modeling assumptions. 

{\it Mean-square angular momentum.}
We construct deformed configurations using the Gogny D1S energy
functional in the HF approximation, constraining on the mass quadrupole operator
$
Q_0 = \langle\Psi|2 z^2 - x^2 - y^2|\Psi\rangle$. 
Here the many-body wave function $\Psi$ is a Slater determinant of orbitals.
It is conventional to characterize the shape by the deformation parameter $\beta$
defined as
\be
\label{beta}
\beta =  \frac{ (5 \pi)^{1/2}}{
3  A^{5/3}r_0^2} Q_0
\ee
where $r_0 = 1.2$ fm.

The general HF formula for the mean-square angular momentum
$\langle J_i^2 \rangle $  around a Cartesian axis $i=x,y,z$ 
is$
\langle \Psi|\hat J^2_i|\Psi \rangle = 
 \sum_{k,k'} n_k (1-n_{k'}) \langle k|
\hat j_i | k'\rangle^2$
where $k,k'$ label a complete set of single-particle orbitals and
$n_k$ (equal to 0 or 1) is the occupation number in the wave
function.   A similar
formula applies to HFB wave functions\cite[Eq. (49)]{alh16},\cite[Eq.
40]{bo07} .

We carry out the constrained minimization of the energy functional
using
the code ${\tt HFBaxial}$ written by one of us (LMR).  
We consider
several daughter nuclei that are prominent among the products of
the $^{235}$U(n,f) reaction, namely
the light fragments \kr, \sr, and \zr, and the heavy fragments
\te, \xe, and \ba.  Since the scission dynamics is still obscure and 
different models can
give very different shapes of the newly-formed fission fragments,  
we
do not attempt to calculate the deformations here but rather 
consider a range.

\begin{figure}[htb] 
\begin{center} 
\includegraphics[width=\columnwidth]{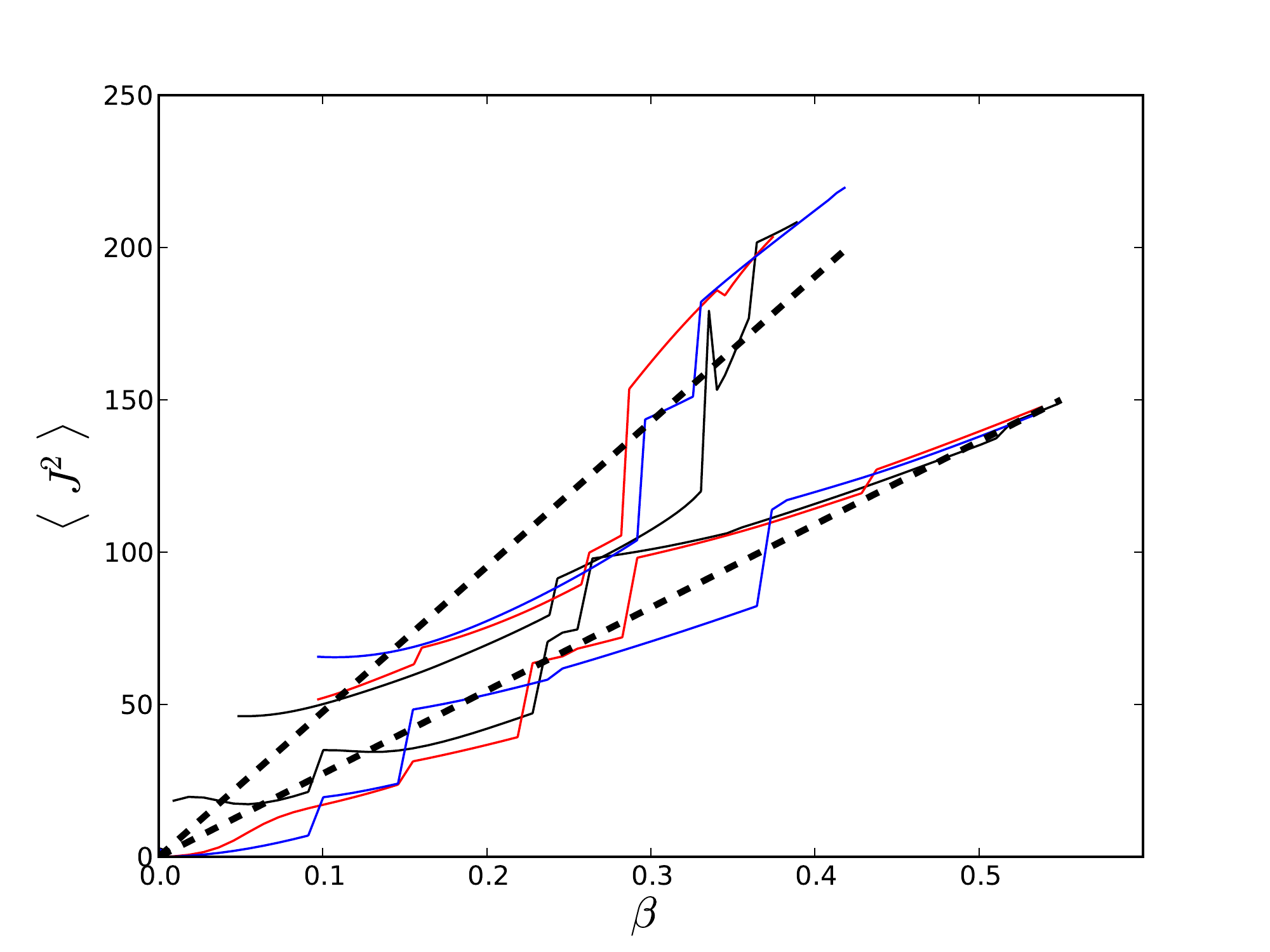}
\caption{Angular momentum of various fission fragments as a function
of deformation $\beta$ (Eq. (\ref{beta}).
}
\label{J2vsb} 
\end{center} 
\end{figure}

The output wave function of ${\tt HFBaxial}$ is
axially symmetric and
invariant under time reversal. 
This implies that the angular
momentum satisfies $\lb J^2_z\rb = 0$ and $\lb J^2_x \rb = \lb J^2_y \rb$.
The code's text output includes 
the expectation value $\langle J_x^2 \rangle$, from which we obtain
$\langle J^2 \rangle = 2\langle J^2_x\rangle$.  
Graphs of $\langle J^2 \rangle_\beta$ are shown in Fig. \ref{J2vsb} 
for the six nuclei mentioned in the previous paragraph.
The curves are far from smooth, due to the strong shell effects
in the HF approximation.    As the deformation increases,
particles jump to orbitals of higher angular momentum, discontinuously
increasing the total.  The dashed lines in the Figure are visual
fits assuming a 
linear relationship.  
A linear formula covering both sets of nuclei is\footnote{Here and
elsewhere the angular momentum is given in units of $\hbar$.}
\be
\label{J2f}
\langle J^2\rangle \approx (0.3 \pm 0.05) A^{3/2} \beta.
\ee

To assess the importance of the deformation contribution we go to
a recent calculation of the $^{239}$Pu(n,f) fission by time-dependent
density functional theory\cite{bu18}.  There it was found that  
the deformation of
the lighter post-scission fragment of mass $A= 105$ was in the
range
$\beta = 0.55-0.7$, depending on the energy functional.  
Applying Eq. (\ref{J2f}), the coherent angular
momentum would be in the range $\langle J^2\rangle = 175 - 220$.  This is much
larger than the estimates $\sim 100$ based on statistical modeling\cite{st14}.
Thus, the coherent angular momentum should be taken into 
account in modeling the de-excitation of the fission products.

\begin{figure}[htb] 
\begin{center} 
\includegraphics[width=\columnwidth]{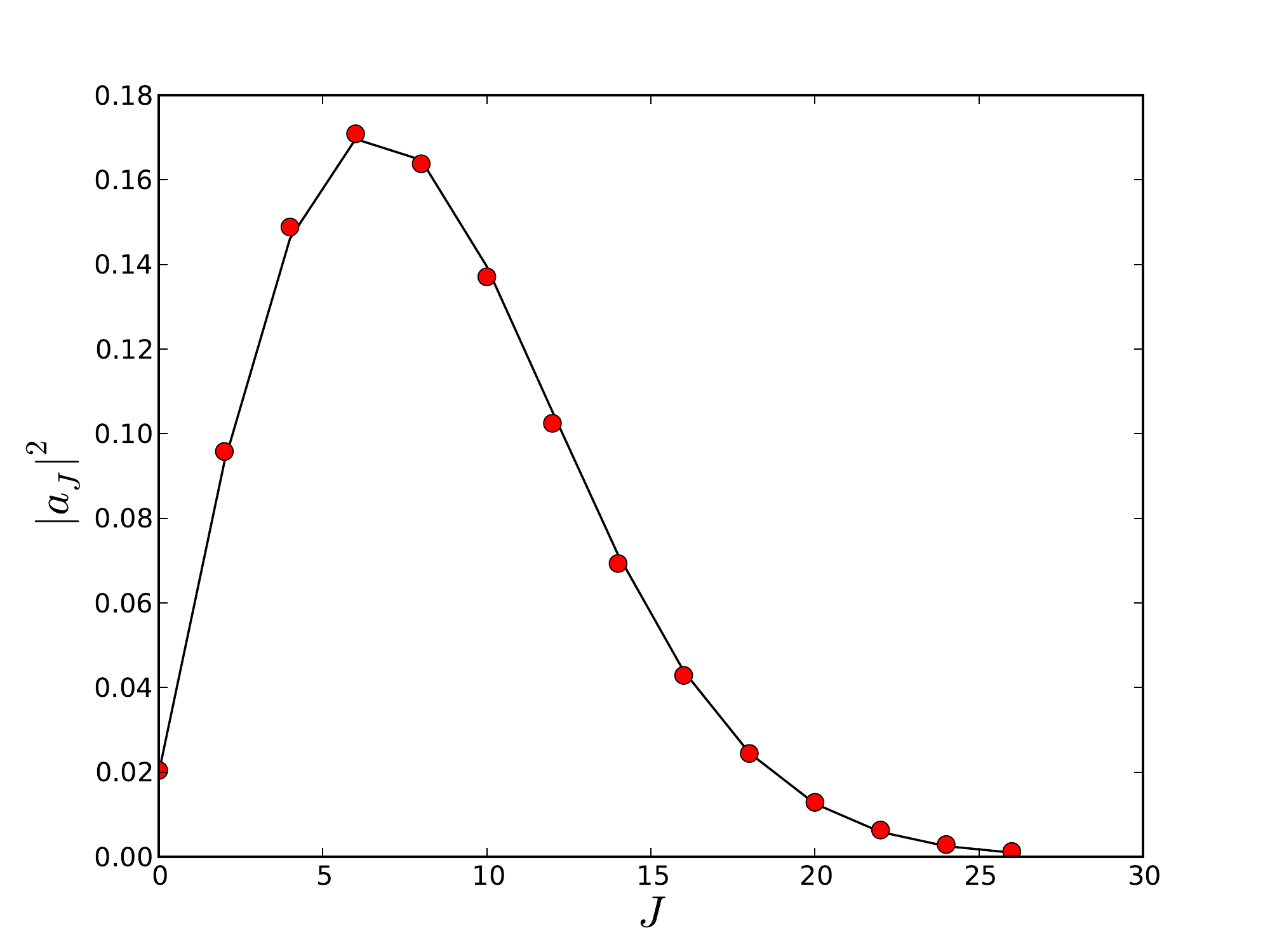}
\caption{ Angular momentum decomposition of a
axially deformed wave function aligned along the
$z$-axis.  Black:  Eq. (\ref{simple2});  Red circles:
Eq. (\ref{simple1}).
\label{aL2}
}
\end{center} 
\end{figure}

{\it J distribution.}  
Next we analyze the distribution of angular
momenta in the deformed wave function.  An aligned axially symmetry
wave function can
be decomposed into angular momentum eigenstates $|JM\rb$ as 
\be
|\Psi\rb = \sum_{J} a_J | J 0 \rb 
\ee 
where $\sum_J |a_J|^2 = 1$, and $J$ is restricted to even
angular momenta for an even-even nucleus in 
its ground state.  
The individual probabilities $|a_J|^2$ can be calculated
by the projection formula 
\be |a_J|^2 = (2 J +1) \int_0^1 d \mu
\lb \Psi |\hat R(\theta) |\Psi\rb P_J(\mu) 
\ee 
where $\hat R$ is
the rotation operator about the $x$ axis, $\mu = \cos \theta$, and $P_J$ is a Legendre polynomial. 
In practice \cite{ha02,be04}, the overlap $\lb \Psi |\hat R(\theta)
|\Psi\rb$ is very well fitted  by an exponential function of
$\mu$.  Then the probabilities $|a_j|^2$ are computed from the
integral
\be 
\label{simple1}
|a_J|^2
= (2 J +1) \int_0^1 d \mu e^{-C(1-\mu^2)} P_J(\mu). 
\ee
Here $C$ is a constant determined from $\langle J^2 \rangle = 
\sum_{J} J (J +1 ) |a_J|^2$; it is approximately given by
$D \approx \langle J^2 \rangle/4$.
The
resulting distribution for mean square angular momentum 
$\lb J_p^2 \rb = \lb
J^2 \rb = 100$ is shown by the red circles in Fig. \ref{aL2}.

In statistical theory, the angular momentum distribution is often parameterized by
Gaussians in the three Cartesian directions,
$
P(J_i)  \sim e^{-J_i^2/2\lb J_i^2 \rb}.
\label{Jdist}
$
Assuming this functional form for the aligned 
intrinsic state and adding a quantum correction, we obtain
the standard spin-cutoff formula \cite[Eq. (3)]{wi72}
\be
|a_J|^2 \sim (J+1/2) e^{-J(J+1)/2\sigma^2}.
\label{simple2}
\ee
where $\sigma^2 \approx \sum_i \lb J^2_i \rb$.
This distribution is shown by the black
line in Fig. \ref{aL2}.  We see that Eq. (\ref{simple2})  is an excellent approximation
to the projection formula Eq. (\ref{simple1}).  

In fact, the same formula Eq.~(\ref{simple2}) emerges from the
semiclassical limit of a Gaussian distribution only in the transverse
directions \cite{bo07}. We note also that Ref. \cite{vo13,ra14} also
assume that the angular momentum is purely transverste.

{\it Statistical decay angular distributions.}
 The gamma-ray angular distribution from fission products 
carries information about the alignment of the deformed fragments.
Indeed, significant anisotropies with
respect to the fission axis were observed a long time ago
\cite{wi72}.  There will also be a component 
due to the Doppler shifts which
we ignore here. 

In the gamma decay $J' \rightarrow J$ the relative
populations of daughter states $\rho(J,M)$
are given in terms of the feeding population distribution $\rho(J',M')$
as 
\be
\label{rhoJM}
\rho(J,M) = \sum_{M'} \rho(J'M') (J'M'L \,M\!\!-\!\!M' | J M)^2.
\ee
Here $L$ is the multipolarity of the electromagnetic transition and
$(J' M'\, L M-M' | J M)$ is a Clebsch-Gordan coefficient.
The angular distribution $p(\theta)$ of the emitted photon is given
by\cite{fe74}
\be
p(\theta) = N_J \sum_{M',M,K} |d^L_{K,1}(\theta)|^2 \rho(J'M')
(J'\, M'\, L\, K | J\,M)^2.
\ee 
where $d^L_{\mu,\mu'}$ is the reduced Wigner $\cal D$-function
and $N_J$ is a normalization constant.

\begin{figure}[htb] 
\begin{center} 
\includegraphics[width=\columnwidth]{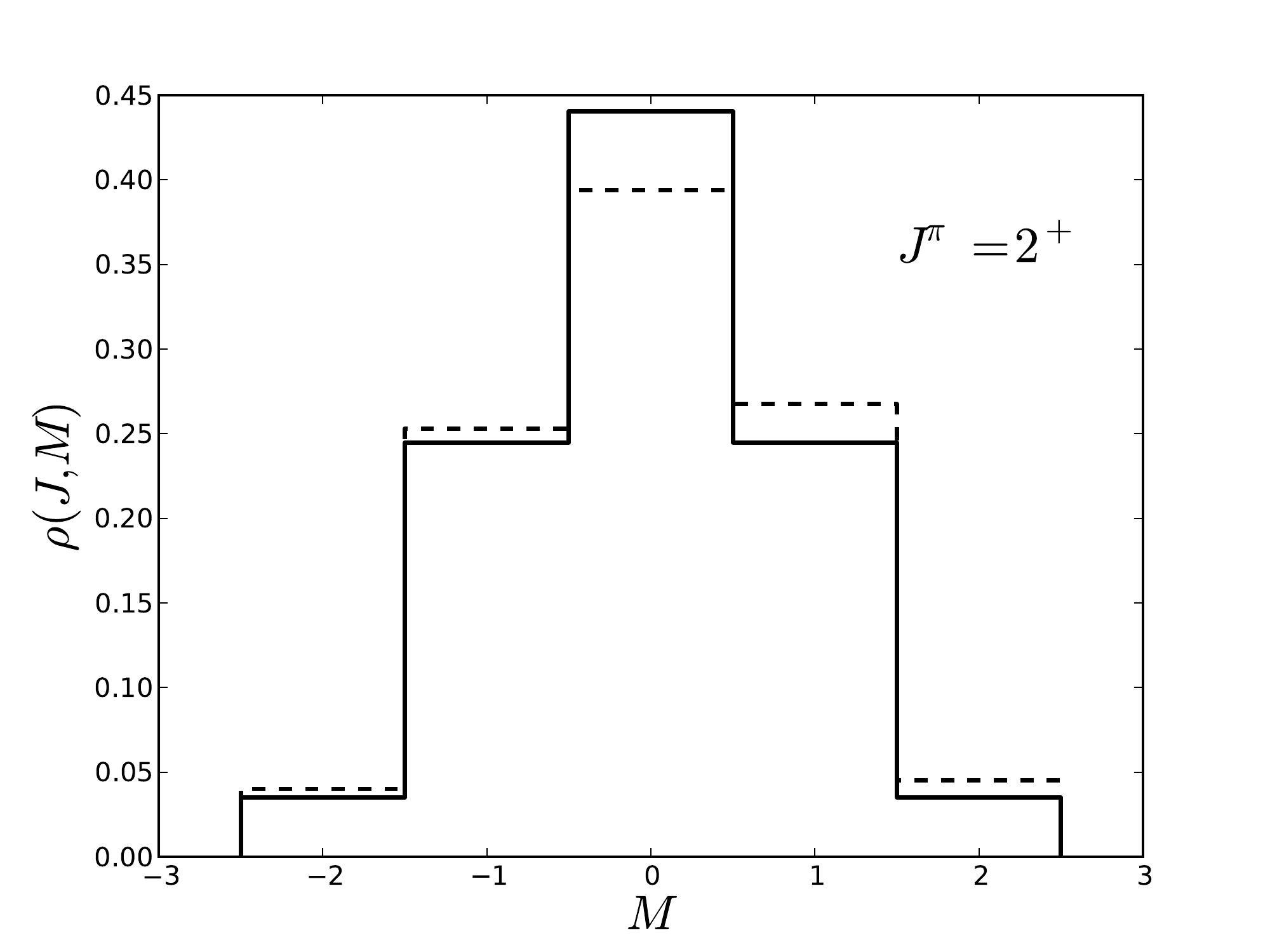}
\includegraphics[width=\columnwidth]{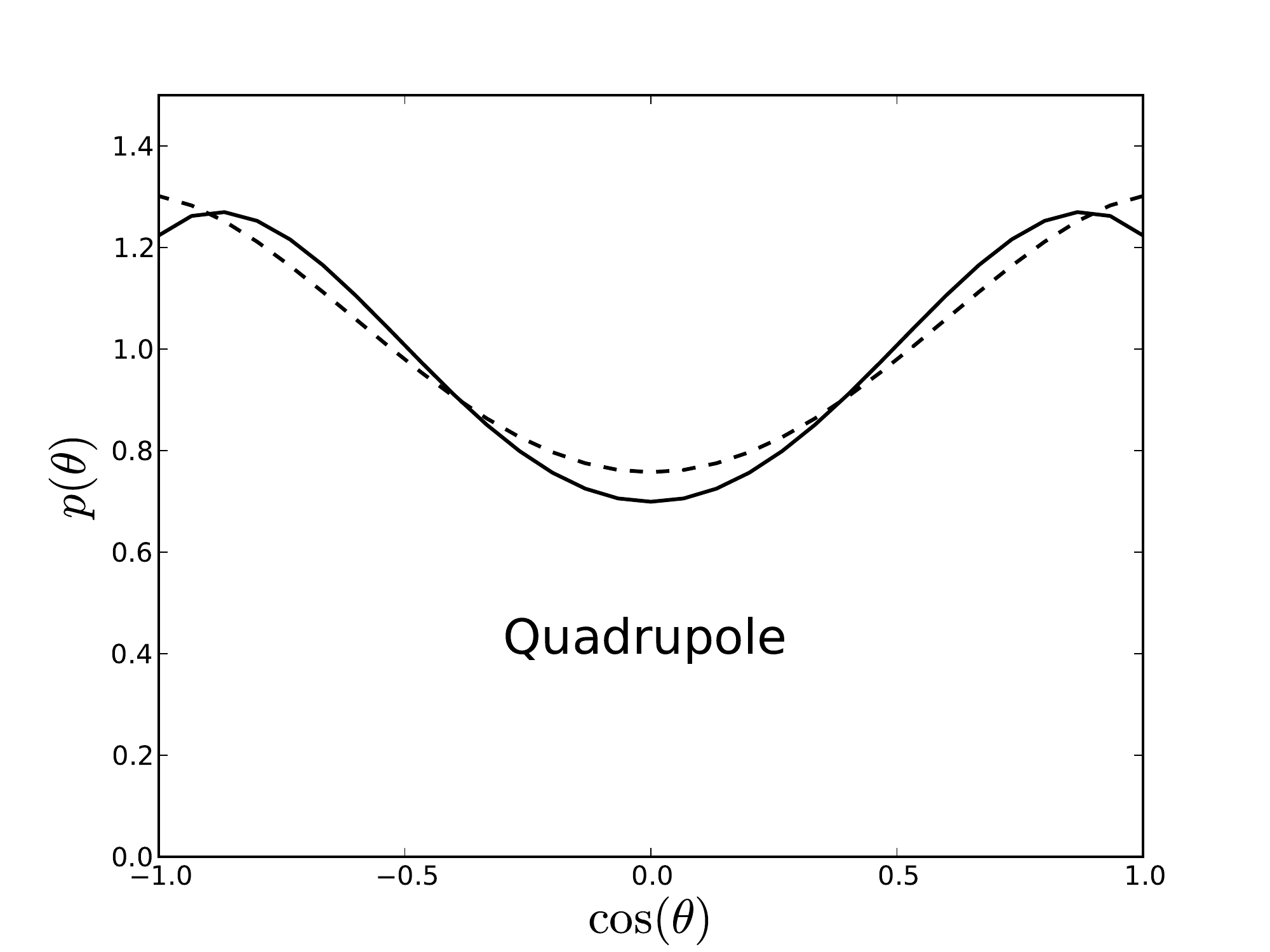}
\caption{
\label{quad}
Top panel:  histogram of $M$-state populations of the $2^+$
first excited level in the cascade.  Solid line is from
the simplified cascade $7 \rightarrow  6\rightarrow 5\rightarrow
4\rightarrow 3\rightarrow 2$ 
starting from the initial distribution
$\rho(7,M) = \delta_{M0}$.  Dashed line is from the realistic
cascade (see text).  Bottom panel:  resulting angular distribution of the
gamma decay $2^+\rightarrow 0^+_{gs}$.
}
\end{center} 
\end{figure}

We first analyze a very simple cascade that starts
from a pure aligned state of angular momentum $(J,M) = (7,0)$. 
The cascade proceeds by emitting dipole photons
until the final transition which is quadrupolar.
Each dipole decay lowers $J$ by one unit until the final quadrupole 
decay.  Thus the decay chain is
$7 \rightarrow  6\rightarrow 5\rightarrow 
4\rightarrow 3\rightarrow 2\rightarrow 0 $.
The population of the $2^+$ first excited state 
remain highly polarized despite the 4-5 preceding gamma decays, as may be seen in 
as the solid line in the upper panel of Fig. \ref{quad}.  
The angular distribution of the subsequent
quadrupolar gamma
ray is shown in the lower panel of the Figure.  It  has an easily 
measurable anisotropy and is peaked along the fission axis.
\begin{figure}[htb] 
\begin{center} 
\includegraphics[width=\columnwidth]{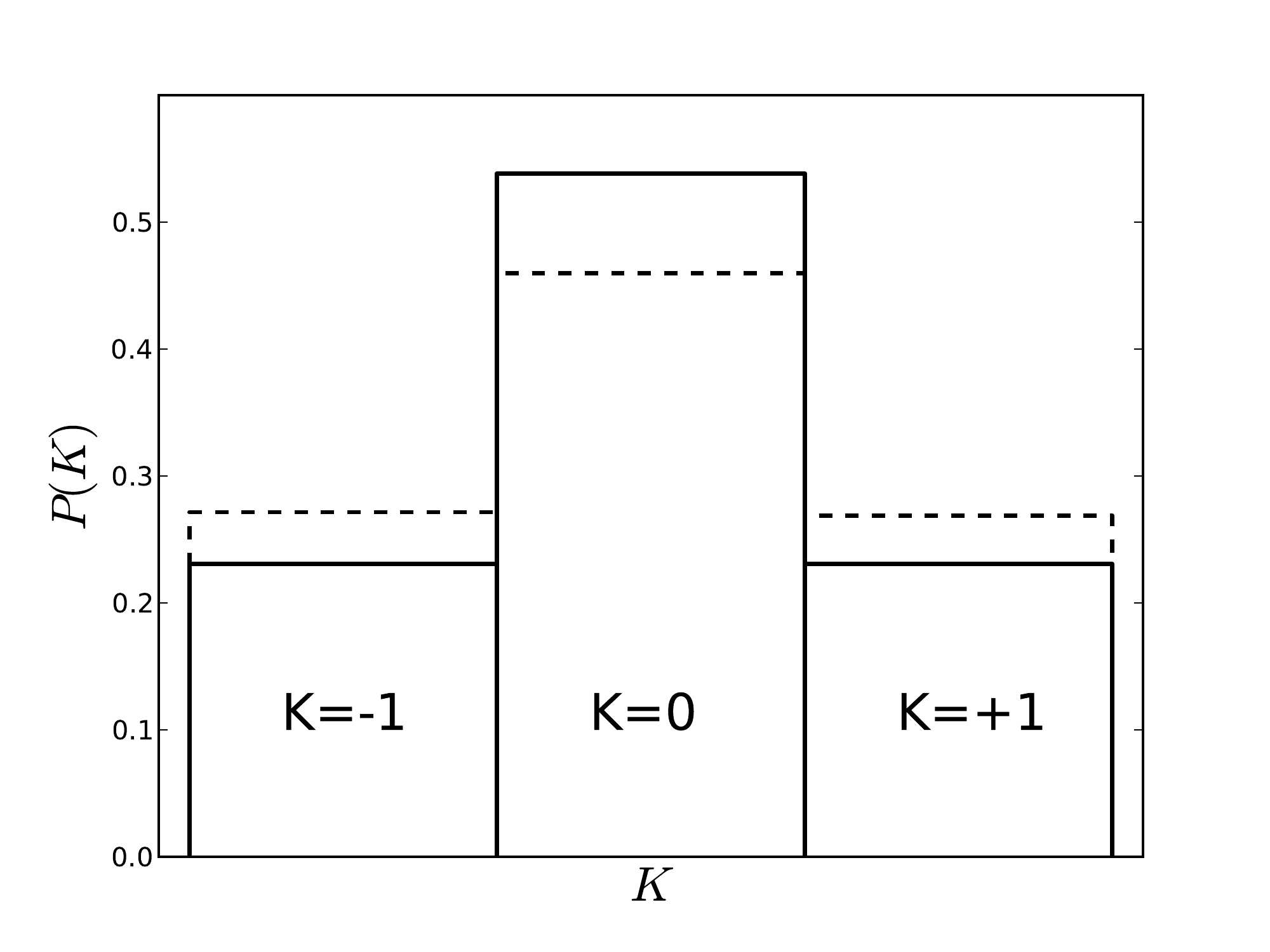}
\includegraphics[width=\columnwidth]{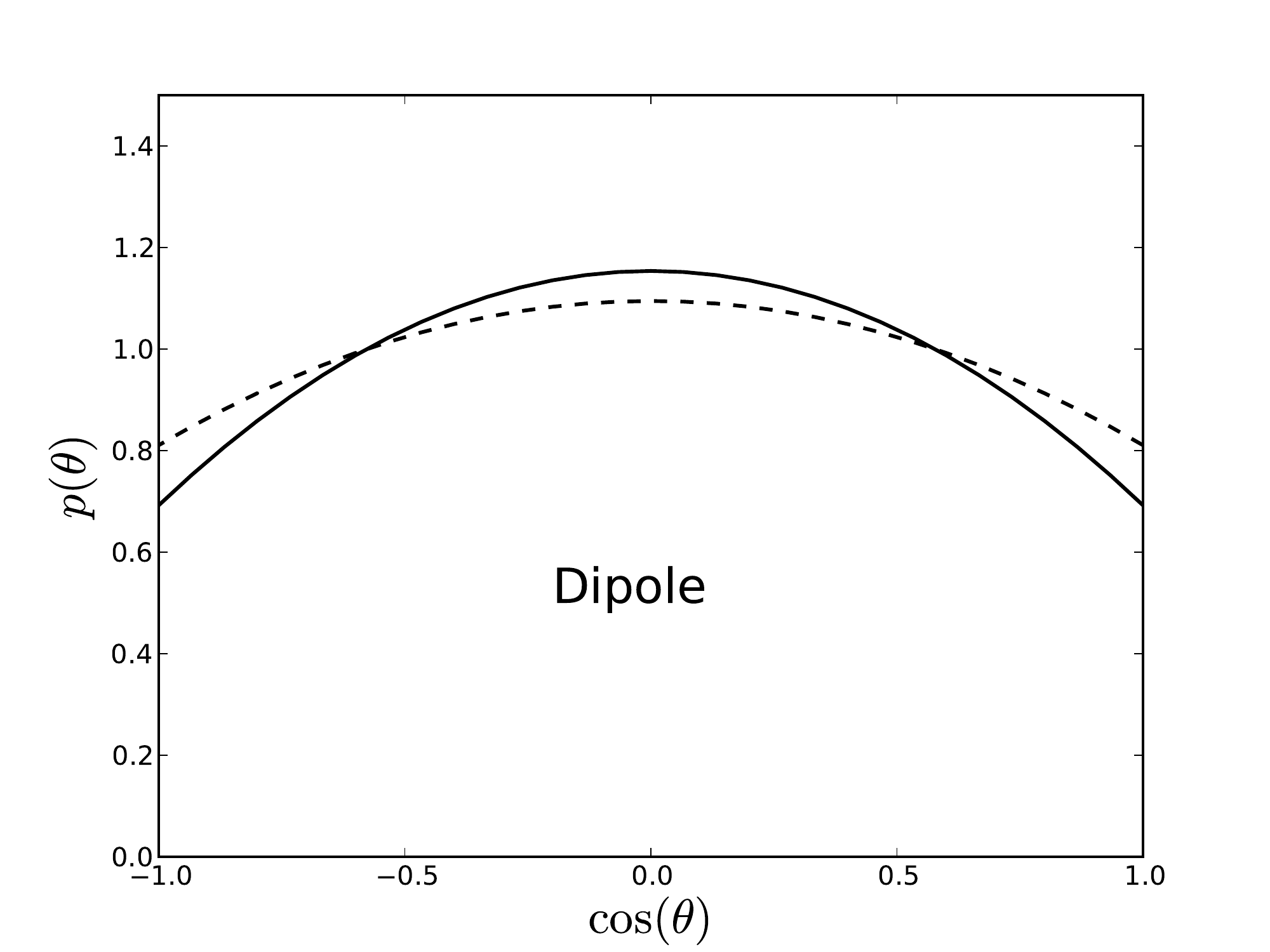}
\caption{
\label{dipole} 
Top panel:  Dipole gamma probabilities as a function of
$K$ of the gamma.  Solid histogram is from the simplified cascade;
dashed line is from the realistic cascade.  Bottom panel:
Dipole gamma angular distribution.
}
\end{center} 
\end{figure}

The dipole photons in the cascade also show an anisotropy.  The
top panel of Fig. \ref{dipole} shows the distribution of dipole
photons with respect to their angular momentum about the fission
axis.  One sees that $M=0$ is favored even though $M=\pm 1$ is
permitted for the great majority of the transitions in the
cascade.  Again the resulting angular distribution (shown in 
the bottom panel of the Figure) is anisotropic,
but now suppressing emission along the fission axis.
 
We have confirmed these findings with a more realistic treatment of 
the level spectrum in the cascade.  The level density is
generated stochastically following the constant-temperature
formula \cite{gi65,be13} $\rho(E^*) = \exp(-E^*/T)/T$ where
$E^*$ is the excitation energy and the parameter $T$ is set to
1 MeV. 
Each level is assigned randomly an angular momentum
and parity $J^\pi$, except as described below.  The probability distribution
for $J$ is given by the spin-cutoff formula Eq. (\ref{simple2}).
Here we assume that the factor $\sigma$ depends on 
excitation energy $E^*$ as \cite{ok18}
$
\sigma = b (E^*)^{1/4} $ with the parameter $b= 4$ MeV$^{-1/4}$.  
The stochastic spectrum is modified in two ways.  First, the
lowest two states are given spin-parity assignments $0^+$
and $2^+$, typical of nearly all even-even nuclei.  We
also limit the maximum $J=J_{max}$ in the probabilistic determination
to insure the cascade will not end on an isomeric state.

The decay branching is also treated stochastically assuming that all
transitions except the final one are electric dipole in character,
with relative transition 
rates given by the
Brink-Axel strength function \cite{ca09}
\be
T_\gamma \propto E^3_\gamma \Gamma \frac{E_\gamma \Gamma}
{(E^2_R- E^2_\gamma)^2 + E^2_\gamma \Gamma^2}.
\ee
The giant resonance parameters are taken as $\Gamma = 5$~MeV
and $E_R = 15$ MeV.  

The angular momentum of the entry point is taken as 
$(J,M) = (8,0)$, chosen to be close to average values obtained from
phenomenological analyses
\cite{na05,li10}.  Its excitation energy should be a little higher
than the neutron separation energy; we set it to
$E^* = 8$ MeV.  Further details are given in the Supplementary
Material \cite{supp}.

The resulting populations of $M$ quantum numbers and gamma
angular distributions are shown as the dashed lines in 
Figs. \ref{quad} and \ref{dipole}.  We see that the qualitative
character of the polarization remain in the more realistic
treatment.  It is common to characterize the anisotropy as 
coefficients of Legendre polynomials.  For the realistic cascade,
we find $p(\theta) = 1 + c_2 P_2(\cos \theta) +...$ with $c_2= -0.2$
and 0.4 for the dipole and quadrupole distributions, respectively.

The qualitative features of the gamma decay angular distribution were 
already seen many years ago \cite{wi72} in a study of the decay products
from the spontaneous fission of $^{252}$Cf.  Measurements were
presented for transitions  from the first excited state to the ground state
in the isotopes $^{144}$Ba, $^{110}$Ru, and $^{105}$Mo; the measured
anisotropy coefficients were $c_2\approx 0.1,0.3,$ and -$0.3$, respectively.  The first
two are electric quadrupole transitions and the signs agree with
expectations, as was indeed noted in the paper.  The higher anisotropy
suggests that heavier
fragment is more spherical, also expected  due to its proximity to the doubly magic
$^{132}$Sn.  The negative $c_2$ for $^{105}$Mo was left unexplained in the
paper.  We now know that spin-parity assignments of the ground and
first excited states in that nucleus \cite{BNL}: the transition is
$7/2^- \rightarrow 5/2^-$ and has
a predominantly M1 character.  Thus, we expect a dipole anisotropy,
as observed.  Overall, the large amplitudes of the measured anisotropies suggest
that quasiparticle excitations do not dominate the
angular momentum distribution of the newly formed fragments.

{\it Outlook}
We hope that the observable discussed here, the prompt gamma angular
distribution, can be used to learn more about the division of excitation
energy in the newly formed fission fragments.  Unfortunately, there are too
many variables to make a direct connection.  The amount of
deformation at scission, and its energy cost, is still very much 
uncertain.  We believe that much of the excitation energy in the
newly formed fission fragments is thermal, in the form of
quasiparticle excitations, but we are still lacking a theory of
the scission process that can describe the sharing of thermal
excitation energy between the two fragments.  
Combining quasiparticle angular momentum with the
deformation will certainly reduce the anisotropy of the gamma radiation, 
and that relationship needs to
be understood quantitatively.  

Another question that needs to be re-examined is the role of
Coulomb excitation in the post-scission acceleration phase.
Both dipole and quadrupole components of the Coulomb field
of the partner fragment are large in first hundred femtoseconds
after scission.  Ref. \cite[Appendix]{wi72} found the effects to
be small in a simple model,  but with present-day theoretical
tools one could make a much more reliable estimate.

{\it Acknowledgments}  
The authors thank 
W. Younes for calling our attention to Ref.
\cite{wi72} and S. Jin for providing deformation parameters from
Ref. \cite{bu18}.  T.K. carried out this work under the auspices of the
National Nuclear Security Administration of the U.S. Department of Energy at Los
Alamos National Laboratory under Contract No. DE-AC52-06NA25396.
The work of LMR was partly supported by Spanish
MINECO grant Nos. FPA2015-65929 and FIS2015-63770.

\end{document}